**Title:** Individuals with Borderline Personality Disorder Show Larger Preferred Social Distance in Live Dyadic Interactions

Authors: Fineberg SK* [a], J Leavitt [a], CD Landry [b], ES Neustadter [c], Lesser, RE [c], DS Stahl [d], S Deutsch-Link [e], PR Corlett [a]

Author affiliations:
[a] Yale University Department of Psychiatry, New Haven, CT
[b] Columbia University College of Physicians and Surgeons, New York, NY
[c] Yale University School of Medicine, New Haven, CT
[d] Knox College, Galesburg, IL
[e] University of Pennsylvania, Philadelphia, PA

Word count (max 4500): 3356
Abstract word count (max 200): 198
Number of tables: 2
Number of figures: 2

Acknowledgements:
We would like to thank Kristin Budde, Carol Gianessi, Megan Ichinose, Taylor McGuiness, Margot Reed, and Erin Feeney for their help as confederates, and Albert Goclowski for his help with preparing and maintaining our testing station at the Connecticut Mental Health Center.

*Sarah Fineberg, MD, PhD
Yale University Department of Psychiatry
Connecticut Mental Health Center
34 Park Street
New Haven CT 06519
Sarah.fineberg@yale.edu
203-974-7265

**Abstract:**

Personal space (PS) regulation is a key component of effective social engagement. PS varies among individuals and is regulated by brain circuits involving the amygdala and the frontoparietal network. Others have reported that simulated PS intrusions suggest larger preferred interpersonal distance (PID) and a central role of amygdala hyperactivity in PS regulation in Borderline Personality Disorder (BPD). This study is the first report of live interpersonal distance preferences and relation to specific symptoms in BPD. We found a 2-fold larger PID in BPD than control (n=30, n=23). There were no significant differences in PID in BPD subject by medication status or pre-study diagnosis, and no significant correlations between PID and intensity of BPD, mood, anxiety, impulsive, or psychotic symptoms. In summary, PID is larger in BPD than control subjects. Unexpectedly, BPD subject PID did not differ in by medication status and did not correlate with intensity of any of the symptom types tested. We discuss these findings in context of severe attachment disturbances in BPD and the relationship between metaphoric social distance in the attachment framework. Future work is needed to identify neural circuits underlying PS regulation in BPD, individual differences in attachment, and relationship to symptom trajectory.

# 1. Introduction

*1.1 Personal Space*

Personal space (PS) refers to "*the area individuals maintain around themselves into which others cannot intrude without arousing discomfort*" and governs each person's preferred interpersonal distance (PID) from others (Hayduk, 1983). PS has been theorized to serve a protective function by regulating one's distance from potential emotional and physical threats while also allowing for an appropriate level of intimacy and trust in social contexts (Lloyd, 2009). For each individual, PID varies according to psychological state and situational circumstances, for example, degree of familiarity with one's interaction partner (Hayduk, 1983), gender roles (Uzzell and Horne, 2006), and emotional valence of the interaction (Tajadura-Jiménez et al., 2011). However, there is also evidence that PS is a stable trait (Perry et al., 2016), varying between individuals according to such factors as attachment style (Yukawa et al., 2007) and varying levels of social anxiety in adults with Autism Spectrum Disorder (Perry et al., 2013).

*1.2 Personal Space: Pathology and Neurobiology*

Changes in personal space regulation have been documented in psychiatric and developmental conditions in which alterations in interpersonal functioning are implicated, including schizophrenia (Holt et al., 2015; Nechamkin et al., 2003; Schoretsanitis et al., 2016; Srivastava and Mandal, 1990), PTSD (Bogovic et al., 2014), claustrophobia (Lourenco et al., 2011), autism spectrum disorders (Asada et al., 2016; Gessaroli et al., 2013; Kennedy and Adolphs, 2014; Perry et al., 2015), and Williams Syndrome (Lough et al., 2016). However, there is limited research on the role of PID and its relationship to disordered attachment or its relationship to Borderline Personality Disorder (BPD).

Kennedy et al. described a patient with bilateral amygdala damage who had profoundly smaller PID compared to matched controls (Kennedy et al., 2009). They also found that intrusion into the control subjects' PS led to amygdala activation (Kennedy et al., 2009). Holt et al. reported activation of dorsal intraparietal sulcus (DIPS) and the ventral premotor cortex (PMv) – two regions previously implicated in near-space processing (Brozzoli et al., 2011) –in healthy control subjects in response to looming photographs of faces (Holt et al., 2014). Furthermore, preferred level of social activity correlated with DIPS-PMv functional connectivity (Holt et al., 2014). Taken together, these findings suggest a crucial role for the fronto-parietal network (DIPS and PMv) and amgydala in regulating PS and PID.

The amygdala is crucially involved in fear conditioning, but is also important for processing emotional cues, such as angry faces (Kim et al., 2016) and ambiguous faces (Davis et al., 2016), for processing non-verbal emotional data such as olfactory fear cues (Hariri and Whalen, 2011), and for social learning (Davis et al., 2010). The fronto-parietal network is implicated in self-regulation (reviewed in (Kelley et al., 2015)) and attentional processes, especially spatial attention (reviewed in (Scolari et al., 2015)), and some argue, internal (self-focused) attention (reviewed in (Luckmann et al., 2014)).

Given that fronto-parietal network dysfunction and amgydala dysregulation are reported in several neuropsychiatric conditions including schizophrenia (Anticevic et al., 2010; Chang et al.,

2014), PTSD (Shin et al., 2006; Weber et al., 2005), and autism (Rudie et al., 2011; Swartz et al., 2013), measures of PID may be subtle markers of neurological abnormalities underlying interpersonal impairments in these conditions.

*1.3 Personal Space and Borderline Personality Disorder*

PS regulation is understudied in borderline personality disorder (BPD), a severe personality disorder characterized by problems with mood, impulse control, and interpersonal functioning (Lieb et al., 2004) as well as altered amygdala function (Schulze et al., 2016), disordered attachment (Levy et al., 2015) and impairments in mentalization (see discussion of mentalization below) (Fonagy et al., 2003).

Disturbed social relationships have been considered central to the phenotype of BPD (Gunderson, 2007) and empirical studies of interpersonal functioning in BPD have found widespread alterations in perceptual biases, social cognition, theory of mind, trust and cooperation (for review see (Lazarus et al., 2014)).

Two recent studies of imagined social experiences have suggested that PS may be altered in BPD. In a study of implicit and explicit behavioral activation, BPD patients and controls were asked how many steps they imagined they would take toward or away from faces shown to them in photographs. BPD patients imagined more steps away from both the happy and fearful faces than controls (Kobeleva et al., 2014). More recently, a neuroimaging study found that simulating personal space intrusion by zooming in on pictures of emotional faces activated fronto-parietal regions and the amygdala in BPD patients and controls (Schienle et al., 2015). BPD patients endorsed a larger PID in a pen and paper task. They also had increased activation of both amygdala and fronto-parietal cortex in the fMRI task, but only towards looming disgusted faces (not faces expressing other emotions). It is important to note, however, that neither of these studies involved live interpersonal interactions – only imagined ones.

Personal space regulation is an essential component of social interaction. However, these dynamic aspects of social exchange have been difficult to capture in traditional experimental probes of social cognition. Typically, studies of PID have relied on derivate stimuli (e.g. disembodied faces) abstracted from social context (Adolphs, 2006; McCall, 2016). When looking at individuals with BPD, utilizing methods such as drawing circles around figures (e.g. Schienle et al., 2016) is likely less reliable than observing live interaction (Harrigan, 2005; Hayduk, 1983; McCall, 2016). Some researchers have argued that individuals with BPD may struggle to accurately describe their emotions in response to hypothetical situations, particularly when they are emotionally aroused (Bateman, 2004). If this is the case, direct observation of behavior may be essential to accurately measuring differences in PID (and other social behaviors) among people with BPD.

*1.4 Attachment, Mentalization, and Borderline Personality Disorder*
Attachment difficulties are hallmarks of many personality disorders and, some argue, are at the root of the interpersonal impairments that are characteristic of most personality disorders (Levy, 2005; Fonagy et al., 1996; Gunderson, 1996; Levy & Blatt, 1999). Attachment theorists posit that bonds between child and primary caregivers early in development have future implications for adult attachment relationships later in life (Bowlby, 1973, 1977). Longitudinal studies have found that attachment patterns evident early in development remain stable over time (Pinquart, 2013). Levy and colleagues theorize that the hypersensitivity to rejection, interpersonal ineffectiveness, and devaluation of the self that are characteristic of BPD can be understood in the context of attachment theory (Levy et al., 2015). Other research has expanded on these findings by demonstrating the relationship between physiological events such as electrodermal activity and heart rate and attachment experiences (Dozier, 1992). A 2006 study indicated that adults with attachment abnormalities demonstrated significant divergence compared to controls between their self-reported reactivity and measured physiological reactivity in response to attachment-related stressors (Diamond, 2006). This finding also argues for the importance of complementing simulated and static social experiments in BPD with data from interactive paradigms.

Mentalizing is the process by which we understand and integrate the feelings and experiences of others, understanding them as separate from ours (Fonagy, 2002). Mentalizing is a distinctly social construct. To effectively mentalize, individuals must have an implicit understanding that their personal experience is subjective and separate from the experience of others (Bateman, 2004). The development of this social cognitive ability is developmental and can be disrupted by factors such as psychological trauma or attachment disturbance during childhood, as have commonly occurred in BPD (Fonagy, 2003). The development of intact mentalization is related to attachment, such that childhood attachment disruption or trauma impairs one's ability to accurately identify the mental states and intentions of others. The phenomenology of BPD may be one consequence of this developmental disruption.

We have previously speculated about the role of embodied cognition and action in BPD (Fineberg et al., 2014), noting the prominence of bodily symptoms and the developmental importance of bodily inputs for making and maintaining attachments. These ideas, taken together with the recent focus on assessing interpersonal functioning in live, interactive contexts (Schilbach, 2016), spurred us to directly test (utilizing face to face, in-vivo interpersonal interactions) preferred interpersonal distance (PID) in adults with BPD. The 2006 finding that individuals with BPD often demonstrate a difference between objective measures of discomfort in attachment-related scenarios and self-reported discomfort, demonstrates the importance of measuring PID with individuals with BPD directly, rather than hypothetically.

To overcome the limitations of previous studies in BPD, we employed a more ecologically valid method. The stop-distance paradigm is a highly reliable measure of PID that is performed in a live 2-person interaction in the lab (Aiello, 1987; Hayduk, 1983; Perry et al., 2016). Study participants are asked to indicate when an approaching confederate has stepped into their personal space. To our knoweldge, this is the first study of personal space regulation in BPD in a live dyadic context, involving face to face interaction with another individual.

**2. Methods:**

*2.1 Subjects:*

Women aged 18-60 were recruited from the community via posters and online advertisements. The study included only women to avoid confounds of differential BPD presentation (e.g. (Mancke et al., 2015)) and differential PID behavior between men and women. This study was approved by the Yale Institutional Review Board and all subjects gave informed consent. Initial screen was done over the phone; subjects were then screened by a psychiatrist (SKF) using semi-structured interviews (Structured Interview for DSM-IV: controls had no psychiatric conditions, BPD subjects had no current substance dependence and no primary psychotic disorder; Revised Diagnostic Interview for Borderline Personality Disorder (Zanarini et al., 2002): controls scored ≤ 4 (scaled total), BPD subjects scored ≥ 8 (scaled total)). Included subjects also read English well (no history of special education, ≤ 11 errors on the Wide Range Achievement Test 4th Edition (WRAT-4) reading test (Wilkinson, 2006)), had intact color vision, had no history of head injury or neurologic condition, and had no gait disturbance.

We collected information on subject education level, hours of work and/or schoolwork per week, current relationship status, and reading level (to detect differences among the more literate subjects in our sample, we used the more challenging North American Adult Reading Test (NAART) instead of the WRAT-4 score here) (Uttl, 2002).

*2.2 Self-report scales:*

Subjects also completed validated self-report scales including the Borderline Symptom List (BSL-23) (Bohus et al., 2009), Beck Anxiety Inventory (BAI) (Beck et al., 1988), Beck Depression Inventory (BDI-II) (Steer et al., 1999), Barratt Impulsiveness Scale (BIS) (Patton et al., 1995), and the Peters Delusion Inventory (PDI) (Peters et al., 2004), including PDI subscales for belief intensity, belief-associated distress, and conviction.

*2.3 Stop distance paradigm:*

To determine PID, subjects began standing face-to-face and 6 feet away from a female confederate. On each of 3 trials, the confederate slowly approached the subject. The confederate was always a young woman, and always stood up straight, made consistent eye contact, and maintained a neutral facial expression. The subjects were instructed that the confederate would walk towards them until they said stop. They were instructed, "Say stop when you feel uncomfortable." Final toe to toe distance was measured using a tape measure, and mean PID for the three trials was computed.

*2.4 Statistics:*

Analyses were conducted using Microsoft Excel and IBM SPSS version 24. Differences in group means were tested using two-tailed Students' t-tests and one-way ANOVAs. We tested for group x trial interactions in the stop distance paradigm with Repeated Measures ANOVA. Correlations were tested using Pearson correlation coefficients.

**3. Results:**

We enrolled 30 women in the control group and 23 women in the BPD group. The two groups were matched on age, years of education, current work status, current relationship status, and reading ability (**Table 1**). Women in the BPD group were significantly more symptomatic on a dimensional measure of BPD symptoms (BSL) as well as self-report measures of depression (BDI), anxiety (BAI), and impulsivity (BIS) (**Table 2**).

Preferred interpersonal distance was more than two-fold larger in the BPD group versus control (**Figure 1**). Repeated-measures ANOVA revealed significant differences by group (F = 16.98, *p*

< 0.001) and by trial (F = 5.21, $p$ = 0.009) but no trial by group interaction (F = 0.196, $p$ = 0.822). Post-hoc tests revealed significant shortening of PID from trial 1 to 2 which was maintained at trial 3 (*Tukey's b* for trial 1 x 2 $p$ = 0.003, for trial 1 x 3 $p$ = 0.052, for trial 2 x 3 $p$ = 0.730). We examined the BPD group for possible differences in PID between subjects based on medication status (**Figure 2A**) or previous mental health diagnosis (**Figure 2B**), however, there were no significant differences between these sub-groups.

We also tested for correlations between preferred interpersonal distance and symptom measures within the BPD group. There were no significant correlations between PID and BSL ($r$ = 0.36, $p$ = 0.56), BDI ($r$ = 0.02, $p$ = 0.93), BAI ($r$ = 0.22, $p$ = 0.36), or BIS ($r$ = 0.26, $p$ = 0.25) score. We also tested for correlations between PID and psychotic-like symptoms, but found no significant correlation to DIB cognitive subscale ($r$ = 0.12, $p$ = 0.64) or PDI scores (total score $r$ = 0.003, $p$ = 0.99, PDI distress $r$ = -0.04, $p$ = 0.89, PDI preoccupation $r$ = -0.07, $p$ = 0.78, PDI conviction $r$ = 0.08, $p$ = 0.77).

## 4. Discussion:

In this study, we found that people with BPD prefer a significantly larger interpersonal distance from a stranger than do non-BPD controls. This finding is not surprising given the existing literature on the role of disordered attachment in BPD and the effects of insecure attachment on social cognition and interpersonal effectiveness. Because attachment was not directly measured as part of our study, we are assuming a link between larger PID and disordered attachment. Further research is needed to confirm this hypothesis. Studies utilizing self-report measures of attachment have consistently found that BPD symptoms are negatively correlated with security of attachment and positively correlated with fearful avoidance attachment (Brennan & Shaver, 1998).

A physiological explanation of our finding that PID is larger in individuals with BPD than in controls would seem to be that the heightened amygdala activity and related negative attribution bias in BPD leads to heightened sensitivity to perceived personal space intrusion. This would fall on the opposite end of the spectrum defined by Kennedy et al. in their study of a patient with bilateral amygdala lesions who tolerated extremely close interpersonal distances without discomfort (Kennedy et al., 2009), and concords with growing data that the amygdala is hyperactive in women with BPD.

In the Schienle *et al.* analysis of BPD subject response to looming faces in fMRI, amygdala activation was observed (Schienle et al., 2015). However, Schulze et al. recently reported that meta-analysis of 19 studies with nearly 300 people in each group revealed increased amygdala activity in response to negative emotional stimuli only in unmedicated, and not in medicated, people with BPD (Schulze et al., 2016). The subjects included in the meta-analysis are similarly aged to our sample, and mostly female, however they are nearly all unmedicated (7/19 studies included unmedicated patients, with 3 studies of all unmedicated patients, 3 with 30-40% unmedicated, and another with 93% unmedicated). Our sample was closer to 50% medicated. To increase homogeneity, the Schulze meta-analysis also excluded studies of interactive social games (a social cooperation investment game, a social rejection game). The included studies delivered negative emotional stimuli such as negative words (3 studies), negative faces (e.g. Reading the Mind in the Eyes Task, 5 studies), negative scenes of people (IAPS, 7 studies), and recalled negative personal experience (4 studies). There were no studies that examined interactive social stimuli (as ours did) included in the meta-analysis. Of note, the Schliene paper examining response to looming faces did observe amygdala activation in both medication and unmedicated patients (64% of the sample was medicated) (Schienle et al., 2015).

Strong conclusions about brain regions likely to be differentially activated in interactive social contexts in BPD are not yet possible given the small literature in this area, but further investigation of amygdala as well as insula and cingulate cortex function in medicated and unmedicated BPD patients, as well as rapidly recovered and unrecovered patients will be critical to understanding the social experience of these patients. This is certainly the case for better understanding of personal space regulation, but also in terms of the regulation of interactive social behavior more generally. The simplicity and ecological validity of the stop-distance paradigm may recommend it as a particularly good assay for further studies in this area.

Personal space regulation and preferred interpersonal distance (in particular) has been more studied in other mental illnesses. Four groups have reported live stop-distance tests in autism spectrum disorder, and results are inconsistent. Asada et al. reported decreased mean PID in 16 men with autism spectrum disorders (ASD), compared to 16 typically developing controls, but similar to our findings, symptom intensity did not significantly correlate to PID (Asada et al., 2016). Previous studies in children with ASD reported conflicting results. However: Gessaroli found increased PID (Gessaroli et al., 2013), whereas Kennedy *et al.* and Perry *et al.* found no group differences. Though in Kennedy *et al.*, parents reported significantly larger numbers of personal space intrusions by their ASD children (Kennedy and Adolphs, 2014) (Perry et al.,

2015). Perry *et al.* did go on to define a potential dimensional symptom measure to explain these conflicting data. Social anxiety, which can vary quite broadly in ASD, correlated with PID in their sample (Perry et al., 2015). They also noted that the N1 ERP signal, an attentional marker which they measured during a personal space simulation game, "significantly and strongly correlated" with IPD within the ASD sample (Perry et al., 2015). We did not find a significant correlation between anxiety symptoms and PID in our BPD subjects, but we did not specifically test social anxiety or attachment style. This will be an important future direction for this work.

Two groups have reported measurements from the stop distance paradigm in schizophrenia, both finding increased PID in schizophrenia versus control (Holt et al., 2015) (Schoretsanitis et al., 2016). One group found that PID correlates to negative, but not positive symptoms (Holt et al., 2015). The other reported that a feeling of threatened paranoia correlated with increased PID, whereas a different flavor of paranoia with a greater sense of personal power abrogated the sensation of personal space intrusion (like the experience of amygdala-lesioned patient). Those patients with schizophrenia but without paranoia behaved on the stop-distance task similarly to control subjects (Schoretsanitis et al., 2016). Holt *et al.* found that greater activation of the fronto-parietal personal space monitoring network in schizophrenia than in controls in response to looming faces vs. withdrawing faces (Holt et al., 2015). Given that many people with BPD experience psychotic-like symptoms, we tested correlations between PID and self-reported measures of psychotic-like symptoms in our sample, but we did not find any significant relationship. This is consistent with the findings above: the self-report data we collected did not specifically distinguish sub-types of paranoid experience.

We identified only one study reporting stop distance behavior in people with post-traumatic stress disorder (PTSD), a condition highly relevant to BPD, as many with BPD report early life trauma, and many symptoms overlap. In a group of 83 male war veterans with chronic PTSD compared to 85 healthy men, PID was increased in PTSD versus control, but as in our sample, the current symptom score did not correlate to PID (Bogovic et al., 2014).

In the mental health field, we lack behavioral assays that report on brain function and prognosis in disorders of social cognition. Future studies should test the stop-distance paradigm as part of an interactive social battery testing state and trait markers of social fear and recovery of healthy social cognition in BPD and mental illness more broadly. Another potential use of PID data may be as a measure of treatment efficacy. Such experiments could test problems with attachment and mentalization: we would expect decreased PID after attachment-focused treatment. To do this, a more direct link would need to be established between disordered attachment and/or difficulties with mentalization and larger PID. This may be accomplished by adding mentalization measures and adult attachment scales (such as the AAI) to this study paradigm in future trials.

**Acknowledgements**

This work was supported by the Connecticut State Department of Mental Health and Addiction Services. Sarah K. Fineberg was supported by NIMH Grant no. 5T32MH019961,"Clinical Neuroscience Research Training in Psychiatry" and a NARSAD Young Investigator Award (2014-2016). Philip R. Corlett was funded by an IMHRO/Janssen Rising Star Translational Research Award and CTSA Grant Number UL1TR000142 from the National Center for Research Resources (NCRR) and the National Center for Advancing Translational Science (NCATS), components of the National Institutes of Health (NIH), and NIH roadmap for Medical Research. Its contents are solely the responsibility of the authors and do not necessarily represent the official view of NIH.

|  | age (yrs) | education (yrs) | work (n- none, pt - <30h, ft - ≥ 30h) | current relationship | reading score (NAART) |
| --- | --- | --- | --- | --- | --- |
| control | 33.4 (13.05) | 15.2 (2.69) | n - 6, pt - 6, ft - 12, missing - 6 | no - 12, yes - 11, missing - 7 | 21.2 (9.04) |
| BPD | 36.9 (12.5) | 14.0 (2.52) | n - 8, pt - 6, ft - 5, missing - 4 | no - 8, yes - 11, missing 4 | 19.77 (7.55) |
| test | t = 0.10 | t = 1.66 | chi sq = 2.62 | chi sq = 0.423 | t = 0.59 |
| p | p = 0.323 | p = 0.1 | p = 0.122 | p = 0.516 | p = 0.55 |

**Table 1.** Subject demographics. Mean results are reported followed by standard deviations in parentheses.

|        | BSL            | BDI            | BAI              | BIS            |
|--------|----------------|----------------|------------------|----------------|
| control | n = 26<br>5.08 (6.2) | n = 27<br>2.56 (4.28) | n = 27<br>6.52 (9.15) | n = 25<br>51.1 (9.59) |
| BPD    | n = 21, 32.19 (19.06) | n = 20, 21.4 (13.19) | n = 20, 23 (12.8) | n = 21, 71.0 (16.5) |
| test   | t = -6.26      | t = -6.15      | t = -4.9         | t = -5.10      |
| p      | $p < 0.001$    | $p < 0.001$    | $p < 0.001$      | $p < 0.001$    |

**Table 2.** Subject symptoms on self-report scales. Mean results of self-report scales are reported followed by standard deviations in parentheses. T-tests reveal significant differences between groups in self-reported Borderline symptoms (BSL), depressive symptoms (BDI), anxiety symptoms (BAI), and impulsivity (BIS).

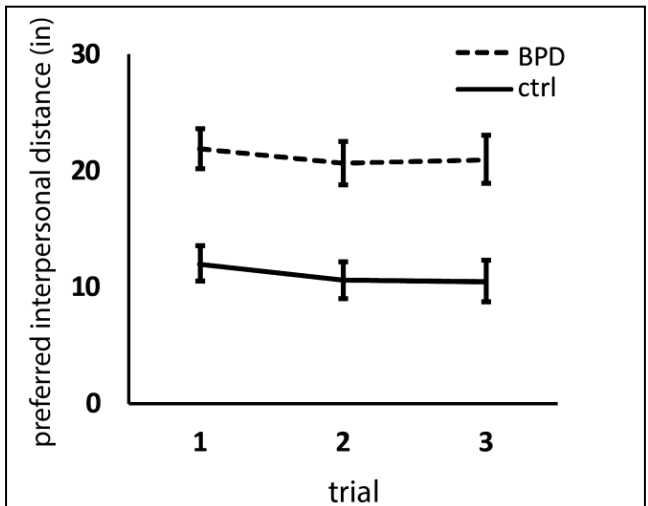

**Figure 1.** Preferred interpersonal distance is significantly greater in BPD than control subjects (Repeated measures ANOVA . F = 16.98, p < 0.001). PID decreases from trial 1 to 2, but not again from trial 2 to trial 3 (F = 5.21, p = 0.009, Tukey's b for trial 1 x 2 p = 0.003, for trial 1 x 3 p = 0.052, for trial 2 x 3 p = 0.730). There was no trial x group interaction (F = 0.196, p = 0.822). Error bars depict standard error of the mean.

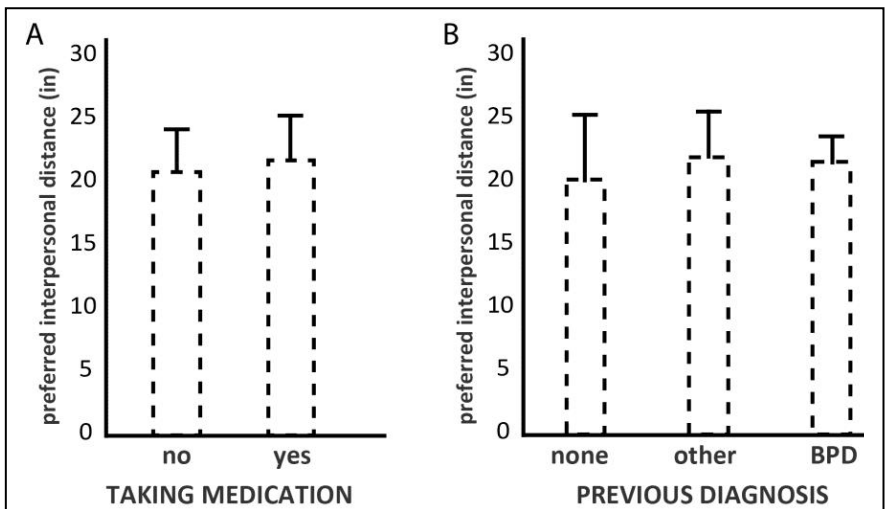

**Figure 2.** Preferred interpersonal distance did not differ between BPD subjects who were **(A)** taking medication or not (Students' t-test t = -0.182, p = 0.86) or BPD subjects who were **(B)** previously undiagnosed, diagnosed with another mental illness, or diagnosed with BPD (ANOVA F = 0.04, p = 0.96). Error bars depict standard error of the mean.